\documentstyle[12pt,epsfig]{article}
\newcommand{\agt}{\,\rlap{\lower 3.5 pt \hbox{$\mathchar \sim$}} \raise 1pt
 \hbox {$>$}\,}
\newcommand{\alt}{\,\rlap{\lower 3.5 pt \hbox{$\mathchar \sim$}} \raise 1pt
 \hbox {$<$}\,}

\catcode`@=11
\newcount\@tempcntc
\def\@citex[#1]#2{\if@filesw\immediate\write\@auxout{\string\citation{#2}}\fi
  \@tempcnta\z@\@tempcntb\m@ne\def\@citea{}\@cite{\@for\@citeb:=#2\do
    {\@ifundefined
       {b@\@citeb}{\@citeo\@tempcntb\m@ne\@citea\def\@citea{,}{\bf ?}\@warning
       {Citation `\@citeb' on page \thepage \space undefined}}%
    {\setbox\z@\hbox{\global\@tempcntc0\csname b@\@citeb\endcsname\relax}%
     \ifnum\@tempcntc=\z@ \@citeo\@tempcntb\m@ne
       \@citea\def\@citea{,}\hbox{\csname b@\@citeb\endcsname}%
     \else
      \advance\@tempcntb\@ne
      \ifnum\@tempcntb=\@tempcntc
      \else\advance\@tempcntb\m@ne\@citeo
      \@tempcnta\@tempcntc\@tempcntb\@tempcntc\fi\fi}}\@citeo}{#1}}
\def\@citeo{\ifnum\@tempcnta>\@tempcntb\else\@citea\def\@citea{,}%
  \ifnum\@tempcnta=\@tempcntb\the\@tempcnta\else
   {\advance\@tempcnta\@ne\ifnum\@tempcnta=\@tempcntb \else \def\@citea{--}\fi
    \advance\@tempcnta\m@ne\the\@tempcnta\@citea\the\@tempcntb}\fi\fi}
\catcode`@=12
\begin{document}
\title{\vskip-3cm{\baselineskip14pt
\centerline{\normalsize DESY 98--012\hfill ISSN~0418--9833}
\centerline{\normalsize MPI/PhT/98--011\hfill}
\centerline{\normalsize hep--ph/9802231\hfill}
\centerline{\normalsize February 1998\hfill}}
\vskip1.5cm
Inclusive $B$-Meson Production in $e^+e^-$ and $p\bar p$ Collisions}
\author{J. Binnewies$^1$, B.A. Kniehl$^2$, and G. Kramer$^1$\\
$^1$ II. Institut f\"ur Theoretische Physik, Universit\"at Hamburg,\\
Luruper Chaussee 149, 22761 Hamburg, Germany\\
$^2$ Max-Planck-Institut f\"ur Physik (Werner-Heisenberg-Institut),\\
F\"ohringer Ring 6, 80805 Munich, Germany}
\date{}
\maketitle
\begin{abstract}
We provide nonperturbative fragmentation functions for $B$ mesons, both at
leading and next-to-leading order in the $\overline{\rm MS}$ factorization
scheme with five massless quark flavors.
They are determined by fitting the fractional energy distribution of $B$ 
mesons inclusively produced in $e^+e^-$ annihilation at CERN LEP1.
Theoretical predictions for the inclusive production of $B$ mesons with high
transverse momenta in $p\bar p$ scattering obtained with these fragmentation 
functions nicely agree, both in shape and normalization, with data recently
taken at the Fermilab Tevatron.

\medskip
\noindent
PACS numbers: 13.60.Le, 13.85.Ni, 13.87.Fh, 14.40.Nd
\end{abstract}
\newpage

\section{Introduction}

The study of $b$-quark production in high-energy hadronic interactions offers
the opportunity to test perturbative quantum chromodynamics (QCD) \cite{1}.
Hadron-collider experiments usually consider the cross section integrated over
a fixed range in rapidity $\eta$ and over all values of transverse momentum
$p_T$ above a variable threshold $p_T^{\rm min}$.
First measurements of such cross sections were performed by the UA1
collaboration at the CERN $p\bar p$ collider with center-of-mass (CM) energy
$\sqrt s=630$~GeV \cite{2}.
More recent experimental results at $\sqrt s=1.8$~TeV were presented by the
CDF \cite{3} and D0 \cite{4} collaborations at the Fermilab Tevatron.
On the theoretical side, such cross sections were calculated up to
next-to-leading order (NLO) in the strong coupling constant $\alpha_s$
\cite{1,5,6}.
The shapes of the theoretical curves agree well with the data of all three
experiments, UA1, CDF, and D0.
However, independent of the beam energy, the absolute normalizations of the
experimental cross sections exceed, by about a factor of two, the respective
predictions obtained with the conventional scale $\mu=\sqrt{p_T^2+m_b^2}$ and
a typical $b$-quark mass of $m_b=4.75$~GeV.
The experimental cross sections could only be reproduced by the theoretical
predictions if $\mu$ and $m_b$ were reduced to $\mu=\sqrt{p_T^2+m_b^2}/2$ and
$m_b=4.5$~GeV and parton density functions (PDF's) with particularly large
values of the asymptotic scale parameter $\Lambda$ were chosen.

In the experimental studies, the $b$-quark production cross section is not 
actually measured directly as a function $p_T^{\rm min}$.
The original measurements refer to the production of $B$ mesons, which decay
either semileptonically, or exclusively or inclusively into $J/\psi$ mesons.
The cross sections for the production of bare $b$ quarks were then obtained
by correcting for the fragmentation of $b$ quarks into $B$ mesons with the 
help of various Monte Carlo (MC) models.
Since this is a model-dependent procedure, it remains unclear whether the
disagreement between the experimental data and the NLO predictions is actually
real.
In order to extract the $b$-quark production cross section, one needs an
independent measurement of the fragmentation of $b$ quarks into $B$ mesons.
In fact, two years ago the CDF collaboration published data on their first
measurement of the $B$-meson differential cross section $d\sigma/dp_T$ for
the exclusive decays $B^+\to J/\psi K^+$ and $B^0\to J/\psi K^{*0}$ based on
the 1992--1993 run (run 1A) \cite{7}.
Similarly to the case of $b$-quark production, the measured cross section was
found to exceed the NLO prediction by approximately a factor of two, while
there was good agreement in the shape of the $p_T$ distribution.
Agreement in the normalization could only be achieved by choosing extreme 
values for the input parameters of the NLO calculation, {\it i.e.}, by 
reducing $\mu$ and $m_b$ and by increasing $\Lambda$ \cite{6}.

In Ref.~\cite{8}, the CDF collaboration extended their analysis \cite{7} by
incorporating the data taken during the 1993--1996 run (run 1B), which yielded
an integrated luminosity of $54.4$~pb$^{-1}$ to be compared with
$19.3$~pb$^{-1}$ collected during run 1A \cite{8}, and presented the cross
section $d\sigma/dp_T$ for the inclusive production of $B^+$ and $B^0$ mesons
with $p_T>6$~GeV in the central rapidity region $|\eta|<1$.
Again, the NLO prediction with input parameters similar to those used for the
integrated $b$-quark cross section in Refs.~\cite{1,5,6}
($\mu=\sqrt{p_T^2+m_b^2}$ and $m_b=4.75$~GeV) was found to agree with the data
in the shape, while its normalization came out significantly too low, by a
factor of $2.1\pm0.4$.
Here, it was assumed that the fragmentations of $b$ quarks into $B$ mesons can
be described by a Peterson fragmentation function (FF) \cite{9} with
$\epsilon=0.006$.
This value for the $\epsilon$ parameter was extracted more than ten years ago
from a global analysis of data on $B$-meson production in $e^+e^-$
annihilation at PEP and PETRA \cite{10}, based on MC models which were then
up-to-date.
The result of this comparison is to be taken with a grain of salt, since the
underlying description of $b\to B$ fragmentation is ad hoc and not backed up
by model-independent data.
It is the purpose of this work to improve on this situation.
The fragmentation of $b$ quarks into $B$ mesons has been measured by the OPAL
collaboration at LEP1 \cite{11}.
The produced $B^+$ and $B^0$ mesons were identified via their semileptonic
decays containing a fully reconstructed charmed meson.
This resulted in the measurement of the distribution of the $B$ mesons in the
scaling variable $x=2E(B)/\sqrt s$, where $E(B)$ is the measured energy of the
$B^+/B^0$ candidate and $\sqrt s=M_Z$ is the LEP1 CM energy.
Earlier measurements of the $b\to B$ FF were reported by the L3 collaboration
at LEP1 \cite{12}.
In the following, we shall base our analysis on the OPAL data, which have 
higher statistics and contain more $x$ bins than the L3 data.

At LEP1, $B$ mesons were dominantly produced by $Z\to b\bar b$ decays with
subsequent fragmentation of the $b$ quarks and antiquarks into $B$ mesons,
which decay weakly.
In the reaction $e^+e^-\to b\bar b\to B+X$ at the $Z$-boson resonance, the
$b$ quarks and antiquarks typically have large momenta.
A large-momentum $b$ quark essentially behaves like a massless particle,
radiating a large amount of its energy in the form of hard, collinear gluons.
This leads to the presence of logarithms of the form
$\alpha_s\ln(M_Z^2/m_b^2)$ originating from collinear singularities in a
scheme, where $m_b$ is taken to be finite.
These terms appear in all orders of perturbation theory and must be resummed.
This can be done by absorbing the $m_b$-dependent logarithms into the FF of
the $b$ quark at some factorization scale of order $M_Z$.
Alternatively, one can start with $m_b=0$ and factorize the collinear
final-state singularities into the FF's according to the $\overline{\rm MS}$
scheme, as is usually done in connection with the fragmentation of light
quarks into light mesons.
This is the so-called massless scheme \cite{13}, in which $m_b$ is neglected,
except in the initial conditions for the FF's.
This scheme was used for NLO calculations of charm and bottom production in
$e^+e^-$ \cite{14}, $p\bar p$ \cite{15}, $\gamma p$ \cite{16,17}, and
$\gamma\gamma $ \cite{18} collisions,
with the additional feature that the massless $c$ and $b$ quarks were
transformed into their massive counterparts by the use of perturbative FF's
\cite{14}.
These perturbative FF's enter as a theoretical input at a low initial scale
$\mu_0$ of order $m_c$ or $m_b$, respectively, and are subject to evolution to
higher scales $\mu$ with the usual Altarelli-Parisi (AP) equations \cite{19}.
Following Ref.~\cite{20}, this theory was extended by including
nonperturbative FF's, which describe the transition from heavy quarks to heavy
mesons \cite{17,21,22}.

In this work, we describe the fragmentation of massless $b$ quarks into $B$
mesons by a one-step process characterized entirely in terms of a
nonperturbative FF, as is usually done for the fragmentation of $u$, $d$, and
$s$ quarks into light mesons.
We assume simple parametrizations of the $b$-quark FF at the starting scale.
We determine the parameters appearing therein through fits to the OPAL data
\cite{11} at lowest order (LO) and NLO.
These $b$-quark FF's are then used to predict the differential cross section
$d\sigma/dp_T$ of $B$-meson production in $p\bar p$ scattering at
$\sqrt s=1.8$~TeV, which can be directly compared with recent data from the
CDF collaboration \cite{8}.

This paper is organized as follows.
In Sect.~2, we recall the theoretical framework for the extraction of FF's
from $e^+e^-$ data, which was previously used for $c$-quark fragmentation into
$D^{*\pm}$ mesons \cite{23}, and present our results for the $b$-quark FF's
at LO and NLO in the $\overline{\rm MS}$ factorization scheme with five 
massless flavors.
We assume three different forms for the FF's at the starting scale, which
enables us to assess the resulting theoretical uncertainty in other kinds of
high-energy processes, such as $p\bar p$ scattering.
In Sec.~3, we apply the nonperturbative FF's thus obtained to predict the
cross section of $B$-meson production in $p\bar p$ collisions at the Tevatron
and compare the result with recent data from CDF \cite{8}.
Our conclusions are summarized in Sec.~4.

\boldmath
\section{$B$-meson production in $e^+e^-$ collisions}
\unboldmath

Our procedure to construct LO and NLO sets of FF's for $B$ mesons is very
similar to the case of $D^{* \pm}$ mesons treated in Refs.~\cite{22,23}.
Here we only give those details which differ from Refs.~\cite{22,23}.

The OPAL data on the inclusive production of $B^+$ and $B^0$ mesons in
$e^+e^-$ annihilation at the $Z$-boson resonance serve as our experimental
input \cite{11}.
These data are presented as differential distributions in $x=2E(B)/\sqrt s$,
where $E(B)$ is the measured energy of the $B^+$ or $B^0$ candidate.
This function peaks at fairly large $x$.
For the fitting procedure we use the experimental $x$ bins, with width
$\Delta x=0.08$, in the interval $0.28<x<1$ and integrate the theoretical
functions over $\Delta x$, which is equivalent to the experimental binning
procedure.
There is a total of nine data points.

When we talk about the $b\to B$ FF, we have in mind the four fragmentation
processes $\bar b\to B^+$, $\bar b\to B^0$, $b\to B^-$, and $b\to \bar B^0$.
In Ref.~\cite{24}, the respective branching fractions are all assumed to be
equal.
If we neglect the influence of the electroweak interactions, this follows from
the $u\leftrightarrow d$ flavor symmetry and the charge-conjugation invariance
of QCD.
We thus make the stronger assumption that the FF's of these four processes all
coincide.
We take the starting scales for the FF's of the gluon and the $u$, $d$, $s$,
$c$, and $b$ quarks and antiquarks into $B$ mesons to be $\mu_0=2m_b$, with
$m_b=5$~GeV.
The FF's of the gluon and the first four quark flavors are assumed to be zero
at the starting scale.
These FF's are generated through the $\mu$ evolution.
For the parametrization of the $b$-quark FF at the starting scale $\mu_0$, we
employ three different forms.
The first one is usually adopted for the FF's of light hadrons, namely
\begin{equation}
\label{standard}
D_b(x,\mu_0)=Nx^\alpha(1-x)^\beta.
\end{equation}
This form has been used in Ref.~\cite{20} to describe the nonperturbative
effects of $b$-quark fragmentation, in addition to a perturbative 
contribution.
The standard (S) parametrization~(\ref{standard}) depends on three free
parameters, $N$, $\alpha $, and $\beta $, which are determined by fits to the
OPAL data \cite{11} after evolution to the factorization scale $M_f=M_Z$.
As our second parameterization, we use the Peterson (P) distribution \cite{9},
\begin{equation}
\label{peterson}
D_b(x,\mu_0)=N\frac{x(1-x)^2}{[(1-x)^2+\epsilon x]^2}.
\end{equation}
This choice is particularly suited to describe a FF that peaks at large $x$.
It has been frequently used in connection with the fragmentation of heavy
quarks, such as $c$ or $b$ quarks, into their mesons.
It depends only on two parameters, $N$ and $\epsilon$.

The third parametrization is theoretically motivated.
There exists a particular class of FF's which are calculable in perturbative 
QCD, namely those of gluons and heavy quarks into heavy-heavy bound states,
such as $c\bar c$, $b\bar b$ \cite{25}, and $c\bar b$ mesons \cite{26}.
These perturbative FF's can also be applied to describe the fragmentation of
$b$ quarks into bound states of $b$ and light quarks, in the sense of a model
assumption rather than a formula derived in perturbative QCD.
The formula for the $\bar b\to B_c$ transition was derived by Braaten (B)
{\it et al.} \cite{26} and reads
\begin{eqnarray}
\label{kingman}
D_b(x,\mu_0)&=&N\frac{rx(1-x)^2}{[1-(1-r)x]^6}
\left[6-18(1-2r)x+(21-74r+68r^2)x^2\right.
\nonumber\\
&&{}-\left.2(1-r)(6-19r+18r^2)x^3+3(1-r)^2(1-2r+2r^2)x^4\right],
\end{eqnarray}
where $r=m_c/(m_b+m_c)$ and $N$ is given in terms of $\alpha_s$, $m_c$, and
the $B_c$-meson wave function at the origin.
Similar formulas also exist for $\bar b\to B_c^*,B_c^{**}$ \cite{26}.
Na\"\i vely applying this formula for $r$ to the fragmentation process
$b\to B$ would yield a rather small number, which is not well determined.
Thus, our philosophy is to treat $N$ and $r$ as free parameters if one of the
quarks in the bound state is light.
In Ref.~\cite{26}, the branching fraction of $c\to B_c$ was found to be two
orders of magnitude smaller than the one of $\bar b\to B_c$.
Extrapolating to the case of $B$ mesons, it hence follows that our assumption
$D_q(x,\mu_0)=0$, where $q$ denotes a light quark, should be well founded even
if $q$ is the light constituent of the $B$ meson.

We calculate the cross section $(1/\sigma_{\rm tot})d\sigma/dx$ for
$e^+e^-\to\gamma,Z\to B^+/B^0+X$ to LO and NLO in the $\overline{\rm MS}$
scheme with five massless quark flavors as described in Ref.~\cite{27}, where
all relevant formulas and references may be found.
In particular, we choose the renormalization and factorization scales to be
$\mu=M_f=\sqrt s$.
As for the asymptotic scale parameter appropriate for five active quark
flavors, we adopt the LO (NLO) value
$\Lambda_{\overline{\rm MS}}^{(5)}=108$~MeV (227~MeV) from Ref.~\cite{27}.
As in Ref.~\cite{22}, we solve the AP equations in $x$ space by iteration of
the corresponding integral equations.
In the Appendix of Ref.~\cite{22}, the timelike splitting functions are listed
in a convenient form, {\it i.e.}, with the coefficients of the delta functions
and plus distributions explicitly displayed.
As in Ref.~\cite{22}, we take the $b$-quark mass to be $m_b=5$~GeV.
Since $m_b$ only enters via the definition of the starting scale $\mu_0$,
its precise value is immaterial for our fit.

The OPAL data are presented in Fig.~3 of Ref.~\cite{11} as the distribution
$dN/dx$ normalized to the bin width $\Delta x=0.08$.
In order to convert these data to the inclusive cross section
$(1/\sigma_{\rm tot})d\sigma/dx$, we need to multiply them by the overall
factor $2R_bf(b\to B)/\Delta x=2.198$, where
$R_b=\Gamma(Z\to b\bar b)/\Gamma(Z\to {\rm hadrons})$, $f(b\to B)$ is the
measured $b\to B$ branching fraction, and the factor of two accounts for the
fact that our cross-section formula \cite{27} includes the fragmentation of
both $b$ and $\bar b$.
Following Ref.~\cite{24}, we identify
$f(b\to B)=f(\bar b\to B^+)=f(\bar b\to B^0)$.
For consistency, we adopt the OPAL results
$R_b=0.2171\pm0.0021\pm0.0021$ \cite{28} and
$f(b\to B)=0.405\pm0.035\pm0.045$ \cite{29}, where the first (second) error
is statistical (systematic).

The values for the input parameters in Eqs.~(\ref{standard}),
(\ref{peterson}), and (\ref{kingman}) which result from our LO and NLO fits to
the OPAL data are summarized in Table~\ref{t1}.
In the following, we refer to these FF's as sets LO~S, NLO~S, LO~P, NLO~P,
LO~B, and NLO~B, respectively.
The corresponding $\chi^2$ values per degree of freedom ($\chi_{\rm DF}^2$)
are listed in the last column of Table~\ref{t1}; there is a total of nine
degrees of freedom.
Except for the sets of type S, the $\chi_{\rm DF}^2$ values for the NLO fits
are slightly lower than those for the LO fits.
The Peterson ansatz~(\ref{peterson}) yields the best fits.
This is surprising, since it has only two free parameters, one less than the
the standard form~(\ref{standard}).
The sets of type B have the largest $\chi_{\rm DF}^2$ values.
Since the $b$-quark FF is peaked at $x\gg0.5$, we have $\alpha\gg\beta$ in the 
case of sets LO~S and NLO~S.
The $\epsilon$ parameters of sets LO~P and NLO~P are larger than the standard
value $\epsilon = 0.006$ \cite{10} usually quoted in the literature.
It is important to note that the values of $\epsilon$ obtained in the various
analyses depend on the underlying theory for the description of the
fragmentation process $b\to B$, in particular, on the choice of the starting
scale $\mu_0$, on whether the analysis is done in LO or NLO (as may be seen
from Table~\ref{t1}), and on how the final-state collinear singularities are
factorized in NLO.
We emphasize that our results for $\epsilon$ in Table~\ref{t1} refer to the
pure $\overline{\rm MS}$ factorization scheme with five massless flavors and
$\mu_0=2m_b=10$~GeV.
If we were to interpret the values for $r$ in Table~\ref{t1} with the formula
$r=m_q/(m_b+m_q)$, which is na\"\i vely adapted from the analogous definition
for $c\bar b$ bound states \cite{26}, then we would find $m_q=688$~MeV and
924~MeV at LO and NLO, respectively.
These values are a factor of 2--3 larger than the generally assumed
constituent-quark masses of the $u$ and $d$ quarks.
This just illustrates the model character of using ansatz~(\ref{kingman}) in
connection with heavy-light bound states.

In Figs.~\ref{f1}(a)--\ref{f1}(c), we compare the OPAL data \cite{11} with the
theoretical results evaluated with sets S, P, and B, respectively. 
Except at low $x$, the LO and NLO results are very similar.
At low $x$, we observe significant differences between LO and NLO.
In this region, the perturbative treatment ceases to be valid.
Here, the massless approximation also looses its validity.
Since $B$ mesons have mass, $m(B)=5.28$~GeV, they can only be produced for
$x>x_{\rm min}=2m(B)/M_Z=0.12$.
The LO result has a minimum in the vicinity of $x_{\rm min}$ and strongly
increases as $x\to0$.
Therefore, our results should only be considered meaningful for
$x\agt x_{\rm cut}$ with $x_{\rm cut}=0.15$, say.
As already observed in connection with the $\chi_{\rm DF}^2$ values, sets
LO~P and NLO~P give the best description of the data.
The contribution due to gluon fragmentation, which only enters at NLO, is
insignificant, below 1\%. 
The contribution due to the first four quark flavors is mostly concentrated
at low $x$ and is also very small.
For $x>x_{\rm cut}$, it makes up less than 1\% of the total integrated cross
section.

It is interesting to study the $b\to B$ branching fraction,
\begin{equation}
\label{branching}
B_b(\mu)=\int_{x_{\rm cut}}^1dx\,D_b(x,\mu),
\end{equation}
where, for reasons explained above, we have introduced a lower cutoff at
$x_{\rm cut}=0.15$.
In Table~\ref{t2}, we present the values of $B_b(\mu)$ at threshold $\mu=2m_b$
and at the $Z$-boson resonance $\mu=M_Z$ for the various FF sets.
As expected, $B_b(\mu)$ is rather stable under the evolution from $2m_b$ to
$M_Z$.
The values of $B_b(M_Z)$ are consistent with the input
$f(b\to B)=0.405\pm0.035\pm0.045$ \cite{29} which was used to scale the
experimental data points \cite{11} so as to obtain the fully normalized cross
section.

Another quantity of interest is the mean $B$ to $b$ momentum fraction,
\begin{equation}
\label{average}
\langle x\rangle_b(\mu)=\frac{1}{B_b(\mu)}\int_{x_{\rm cut}}^1dx\,x
D_b(x,\mu).
\end{equation}
Table~\ref{t2} also contains the values of $\langle x\rangle_b(\mu)$ at
$\mu=2m_b$ and $M_Z$ evaluated with the various FF sets.
The differences between sets S, P, and B on the one side and between LO and
NLO on the other side are small.
As $\mu$ runs from $2m_b$ to $M_Z$, $\langle x\rangle_b(\mu)$ decreases from
approximately 0.8 down to below 0.7.
This is a typical feature of the $\mu$ evolution, which generally softens the
FF's.
Our values of $\langle x\rangle_b(M_Z)$ can be compared with the experimental
result reported by OPAL \cite{11},
\begin{equation}
\label{opal}
\langle x\rangle_b(M_Z)=0.695\pm0.006\pm0.003\pm 0.007,
\end{equation} 
where the errors are statistical, systematic, and due to model dependence,
respectively.
Our results in Table~\ref{t2} are in reasonable agreement with
Eq.~(\ref{opal}).
In connection to this, we remark that Eq.~(\ref{opal}) is not directly
obtained from the measured distribution, which would be difficult to do, since
there are no data points below $x=0.2$.
To extrapolate to the unmeasured region, OPAL uses four different models which
describe the primordial fragmentation of $b$ quarks inside their MC
simulation.
Equation~(\ref{opal}) is actually determined from the MC fits to the measured
data points.
Obviously, the quoted error for the model dependence can only account for the
specific model dependence inside their particular MC approach, and need not be
characteristic of the absolute model dependence.
A rather model-independent fit to the $x$ distribution, including a MC
estimate for the region $x<0.2$, leads to
$\langle x\rangle_b(M_Z)=0.72\pm 0.05$ \cite{11}, where the error is only
statistical and does not account for the uncertainty due to the extrapolation.
Our results in Table~\ref{t2} are somewhat smaller than this value and are
barely consistent with the experimental error given above.
Nevertheless, we believe that our results in Table~\ref{t2} are in reasonable
agreement with the independent determinations of $\langle x\rangle_b(M_Z)$
quoted in Ref.~\cite{11}.

\boldmath
\section{$B$-meson production in $p\bar p$ collisions}
\unboldmath

In this section, we compare our LO and NLO predictions for the cross section
of inclusive $B^+/B^0$ production in $p\bar p$ collisions at the Tevatron
($\sqrt s=1.8$~TeV) with recent data from the CDF collaboration \cite{8}.
These data come as the $p_T$ distribution $d\sigma/dp_T$ integrated over the
central rapidity region $|\eta|<1$ for $p_T$ values between 7.4 and 20 GeV.
They are normalized in such a way that they refer to the single channel
$p\bar p\to B^++X$.
In the case of run 1A, where both $B^+$ and $B^0$ mesons were detected, the
respective cross sections were averaged, {\it i.e.}, their sum was divided by
a factor of two.

Our formalism is very similar to Ref.~\cite{30}, where inclusive light-meson
production in $p\bar p$ collisions was studied in the QCD-improved parton
model.
The relevant formulas and references may be found in Ref.~\cite{30}, and we 
refrain from repeating them here.
We work at NLO in the $\overline{\rm MS}$ scheme with $n_f=5$ massless
flavors.
For the proton and antiproton PDF's, we use set CTEQ4M \cite{31} with
$\Lambda_{\overline{\rm MS}}^{(5)}=202$~MeV.
We evaluate $\alpha_s$ from the two-loop formula with this value of
$\Lambda_{\overline{\rm MS}}^{(5)}$.
We recall that the evolution of the FF sets NLO~S, NLO~P, and NLO~B is
performed with $\Lambda_{\overline{\rm MS}}^{(5)}=227$~MeV, which is very
close to the above value.
We identify the factorization scales associated with the proton, antiproton,
and the $B$ meson and collectively denote them by $M_f$.
We choose renormalization and factorization scales to be $\mu=M_f=2m_T$, where
$m_T=\sqrt{p_T^2+m_b^2}$ is the $B$-meson transverse mass.
Whenever we present LO results, they are consistently computed using set
CTEQ4L \cite{31} of the proton and antiproton PDF's, our LO sets of $B$-meson 
FF's, the one-loop formula for $\alpha_s$ with
$\Lambda_{\overline{\rm MS}}^{(5)}=181$~MeV \cite{31}, and the LO
hard-scattering cross sections.
We adopt the kinematic conditions from Ref.~\cite{8}.
Since we employ $D_b(x,\mu)$ both for the $b$ and $\bar b$ quarks, the 
resulting cross section corresponds to the sum of the $B^+$ and $B^-$ yields.
Thus, it needs to multiplied by a factor of 1/2, in order to match the cross 
section quoted in Ref.~\cite{8}.

First, we consider the $p_T$ distribution $d\sigma/dp_T$ integrated over 
the rapidity region $|\eta|<1$ as in the CDF analysis \cite{8}.
In Fig.~\ref{f2}(a), we compare the CDF data \cite{8} with the LO and NLO
predictions evaluated with our various sets of $B$-meson FF's.
The NLO distributions fall off slightly less strongly with increasing $p_T$
than the LO ones.
The results for sets LO~S, LO~P, and LO~B almost coincide.
The same is true of the results for sets NLO~S, NLO~P, and NLO~B.
This means that the details of the $b\to B$ fragmentation is tightly
constrained by the LEP data, and that the considered variation in the
functional form of the $b$-quark FF at the starting scale has very little
influence on the $p_T$ distribution.
Henceforth, we shall only employ sets LO~P and NLO~P, which yielded the best
fits to the OPAL data \cite{11}.
We observe that our prediction agrees very well with the CDF data, within
their errors.
This is even true for the data point with smallest $p_T$, $p_T=7.4$~GeV,
where the massless approach is presumably not valid any more.
It should be emphasized that the NLO prediction reproduces both the shape and
the absolute normalization of the measured $p_T$ distribution, while the
previous investigations mentioned in the Introduction fell short of the data
by roughly a factor of two.

The CDF collaboration has not yet presented results on the $\eta$ distribution
of the produced $B$ mesons, which would allow for another meaningful test of
the QCD-improved parton model endowed with $B$-meson FF's.
Anticipating that such a measurement will be done in the future, we show in
Fig.~\ref{f2}(b) the $\eta$ dependence of $d^2\sigma/d\eta dp_T$ evaluated
with sets LO~P and NLO~P at $p_T=13.4$, 17.2, 20, and 30~GeV.
The first three of these $p_T$ values are among those for which CDF performed 
measurements of $d\sigma /dp_T$ \cite{8}.
Since the $\eta$ spectrum is symmetric around $\eta=0$, we only consider
$\eta\ge0$ in Fig.~\ref{f2}(b).
As expected, the cross section falls off with $\eta$ increasing from zero up 
to the kinematic limit, which depends on $p_T$.

In order to assess the reliability of our predictions, at least to some
extent, we now investigate the scale dependence of the cross section
considered in Fig.~\ref{f2}(a).
To this end, we introduce the scale factor $\xi$ such that $\mu=M_f=2\xi m_T$.
In Fig.~\ref{f3}, the $\xi$ dependence of $d\sigma/dp_T$ is displayed for
$p_T=13.4$, 20, and 30~GeV.
The calculation is performed with sets LO~P and NLO~P.
For the two highest $p_T$ values, $p_T=20$ and 30~GeV, we observe the expected
pattern.
The LO results for $d\sigma/dp_T$ essentially decrease with $\xi$ increasing,
whereas the NLO results are rather $\xi$ independent and exhibit points of
horizontal tangent close to $\xi=1$.
Furthermore, the LO and NLO curves intersect near these points.
Thus, the scale choice $\xi=1$ is favoured both from the principles of minimal
sensitivity \cite{32} and fastest apparent convergence \cite{33}.
These observations reassure us of the perturbative stability and the
theoretical soundness of our calculation in the upper $p_T$ range.
For $p_T=13.4$~GeV, the NLO prediction of $d\sigma/dp_T$ shows a stronger
scale dependence, in particular, when the scale is drastically reduced.
If we limit the scale variation to the interval $1/2<\xi<2$, which is 
frequently considered in the literature, the NLO cross section still varies
by a factor of 1.56, to be compared with 1.15 at $p_T=20$~GeV.
We hence conclude that, below $p_T=13.4$~GeV, our NLO predictions should be
taken with a grain of salt.
The dents in the curves for $p_T=13.4$~GeV appear at the value of $\xi$ where
$M_f=\mu_0$.
This is because we identify $D_b(x,M_f)=D_b(x,\mu_0)$ if $M_f<\mu_0$,
{\it i.e.}, the FF's are frozen below their threshold.

We must also remember that, for $p_T$ values comparable to $m_b$, the 
massless-quark approximation ceases to be valid, since terms of order
$m_b^2/p_T^2$ are then not negligible anymore.
For $p_T=13.4$~GeV and 20~GeV, we have $m_b^2/p_T^2=0.14$ and 0.063,
respectively, so that the massless approximation should certainly be valid for
$p_T=20$~GeV.
On the other hand, for $p_T=20$~GeV, we have $\alpha_s\ln(p_T^2/m_b^2)=0.42$,
assuming that $\alpha_s=0.15$, so that the NLO calculation in the massive
scheme, where these logarithmic terms are not resummed, should then already be 
inadequate.
From these considerations, we conclude that our predictions should be fairly
reliable for $p_T\agt15$~GeV.

\section{Conclusions}

In this paper, we considered the inclusive production of single $B$ mesons in
the QCD-improved parton model endowed with nonperturbative FF's.
We chose to work at NLO in the pure $\overline{\rm MS}$ factorization scheme
with five massless quark flavors.
This theoretical framework is known to lead to an excellent description of a
wealth of experimental information on inclusive light-hadron production in
different types of high-energy reactions \cite{27,34}.
It is thus expected to also work well in the case of $B$ mesons provided that
the characteristic mass scale $M$ of the process by which they are produced is
large compared to the $b$-quark mass.
Then, the large logarithms of the type $\alpha_s\ln(M^2/m_b^2)$ which are 
bound to arise in any scheme where bottom is treated as a massive flavor get
properly resummed by the AP evolution, while the omission of the terms
suppressed by powers of $m_b^2/M^2$ is a useful approximation.
The criterion $M\gg m_b$ is certainly satisfied for $e^+e^-$ annihilation on
the $Z$-boson resonance, and for hadroproduction of $B$ mesons with
$p_T\gg m_b$.
Owing to the factorization theorem, the FF's are universal in the sense that 
they just depend on the produced hadrons and the partons from which they
sprang, but not on the processes by which the latter were produced.
Thus, the theoretical framework adopted here is particularly suited for a
consistent description of LEP1 and high-$p_T$ Tevatron data of inclusive
$B$-meson production.
By the same token, a massive calculation at fixed order would be
inappropriate for this purpose.

Our procedure was as follows.
We determined LO and NLO $B$-meson FF's by fitting the fractional energy
distribution of the $B$-meson sample collected by the OPAL collaboration at
LEP1 \cite{11}.
In order to get some handle on the theoretical uncertainty, we adopted three
different functional forms for the $b\to B$ FF at the starting scale, which we
took to be $\mu_0=2m_b=10$~GeV.
The ansatz proposed by Peterson {\it et al.}\ \cite{9} yielded the best LO and
NLO fits, with $\chi_{\rm DF}^2=0.67$ and 0.27, respectively.
The $\epsilon$ parameter, which measures the smearing of the Peterson 
distribution, came out as 0.0126 and 0.0198, respectively, {\it i.e.}, more 
than twice as large as the standard value $\epsilon=0.006$ determined by
Chrin \cite{10} more than a decade ago, before the LEP1 era.
In this connection, we should emphasize that the results for the fit 
parameters, including the value of $\epsilon$, are highly scheme dependent at
NLO, and must not be na\"\i vely compared disregarding the theoretical
framework to which they refer.
The $b\to B$ branching fraction and the mean $B$ to $b$ momentum fraction
evaluated from the resulting FF's after the evolution to $\mu=M_Z$ turned out
to be in reasonable agreement with the model-dependent determinations by OPAL
\cite{11}.
Using our FF's, we made theoretical predictions for the inclusive
hadroproduction of single $B$ mesons with large $p_T$.
We found good agreement, both in shape and normalization, with the $p_T$
distribution recently measured in the central rapidity region by the CDF
collaboration at Fermilab \cite{8}.
From the study of the scale dependence of the LO and NLO calculations, we
concluded that our results should be reliable for $p_T\agt15$~GeV.
To our surprise, the central prediction, with scales $\mu=M_f=2m_T$, also
nicely agreed with the CDF data in the low-$p_T$ range, where the massless
scheme is expected to break down.
We recall that the massive NLO calculation with traditional Peterson
fragmentation \cite{10} was found to fall short of these data by a factor of
two.
It would be interesting to also test the predicted $\eta$ distribution against
experimental data.

\bigskip
\centerline{\bf ACKNOWLEDGMENTS}
\smallskip\noindent
We are grateful to Ties Behnke and Alex Martin for clarifying communications
concerning Ref.~\cite{11}.
One of us (G.K.) thanks the Theory Group of the Werner-Heisenberg-Institut for
the hospitality extended to them during a visit when this paper was prepared.
The II. Institut f\"ur Theoretische Physik is supported by Bundesministerium
f\"ur Bildung und Forschung, Bonn, Germany, under Contract 05~7~HH~92P~(0),
and by EU Program {\it Human Capital and Mobility} through Network
{\it Physics at High Energy Colliders} under Contract
CHRX--CT93--0357 (DG12 COMA).

\newpage

\begin{table}
\centerline{\bf TABLE CAPTIONS}
\bigskip

\caption{Fit parameters for the $b\to B$ FF's according to sets S, P, and B at
LO and NLO and respective values of $\chi^2$ per degree of freedom.
All other FF's are taken to be zero at the starting scale $\mu_0=2m_b=10$~GeV.
\protect\label{t1}}

\caption{$b\to B$ branching fractions and mean $B$ to $b$ momentum fractions
evaluated from Eqs.~(\protect\ref{branching}) and (\protect\ref{average}),
respectively, at the starting scale and at the $Z$-boson resonance using the
various FF sets.
\protect\label{t2}}

\end{table}

\newpage

\begin{figure}
\centerline{\bf FIGURE CAPTIONS}
\bigskip

\caption{The cross sections of inclusive $B^+/B^0$-meson production in
$e^+e^-$ annihilation at $\protect\sqrt s=M_Z$ evaluated with sets (a) LO~S
and NLO~S, (b) LO~P and NLO~P, and (c) LO~B and NLO~B are compared with the
OPAL data \protect\cite{11}.
\protect\label{f1}}

\caption{(a) The cross section $d\sigma/dp_T$ of inclusive $B^+/B^0$-meson
production in $p\bar p$ collisions with $\protect\sqrt s=1.8$~TeV, integrated
over $|\eta|<1$, is compared with the CDF data \protect\cite{8}.
The predictions are calculated at LO and NLO with sets S, P, and B.
(b) The cross section $d\sigma/d\eta dp_T$ at fixed values of $p_T$ evaluated
with sets LO~P and NLO~P.
\protect\label{f2}}

\caption{Scale dependence of the cross section $d\sigma/dp_T$, integrated over
$|\eta|<1$, at fixed values of $p_T$.
The predictions are calculated at LO and NLO with set P.
\protect\label{f3}}

\end{figure}

\newpage

\begin{table}
\begin{center}
\begin{tabular}{|c|c|c|c|c|c|c|}
\hline
set & $N$ & $\alpha$ & $\beta$ & $\epsilon$ & $r$ & $\chi_{\rm DF}^2$ \\
\hline
\hline
LO S & 56.4 & 8.39 & 1.16 & -- & -- & 0.80 \\
\hline
NLO S & 79.4 & 8.06 & 1.45 & -- & -- & 1.21 \\
\hline
\hline
LO P & 0.0952 & -- & -- & 0.0126 & -- & 0.67 \\
\hline 
NLO P & 0.116 & -- & -- & 0.0198 & -- & 0.27 \\
\hline 
\hline
LO B & 0.308 & -- & -- & -- & 0.121 & 2.50 \\
\hline 
NLO B & 0.280 & -- & -- & -- & 0.156 & 1.66 \\
\hline 
\end{tabular}

\smallskip
{\bf Table~1}

\vspace{1cm}

\begin{tabular}{|c|c|c|c|c|}
\hline
set & $B_b(2m_b)$ & $B_b(M_Z)$ & $\langle x\rangle_b(2m_b)$ &
$\langle x\rangle_b(M_Z)$ \\
\hline 
\hline
LO S  & 0.425 & 0.411 & 0.813 & 0.697 \\
\hline 
NLO S & 0.384 & 0.370 & 0.787 & 0.672 \\
\hline
\hline
LO P  & 0.448 & 0.431 & 0.787 & 0.677 \\
\hline
NLO P & 0.405 & 0.388 & 0.758 & 0.650 \\
\hline
\hline
LO B  & 0.460 & 0.442 & 0.768 & 0.663 \\
\hline
NLO B & 0.416 & 0.398 & 0.739 & 0.635 \\
\hline
\end{tabular}

\smallskip
{\bf Table~2}

\end {center}
\end{table}

\newpage
\begin{figure}[ht]
\epsfig{figure=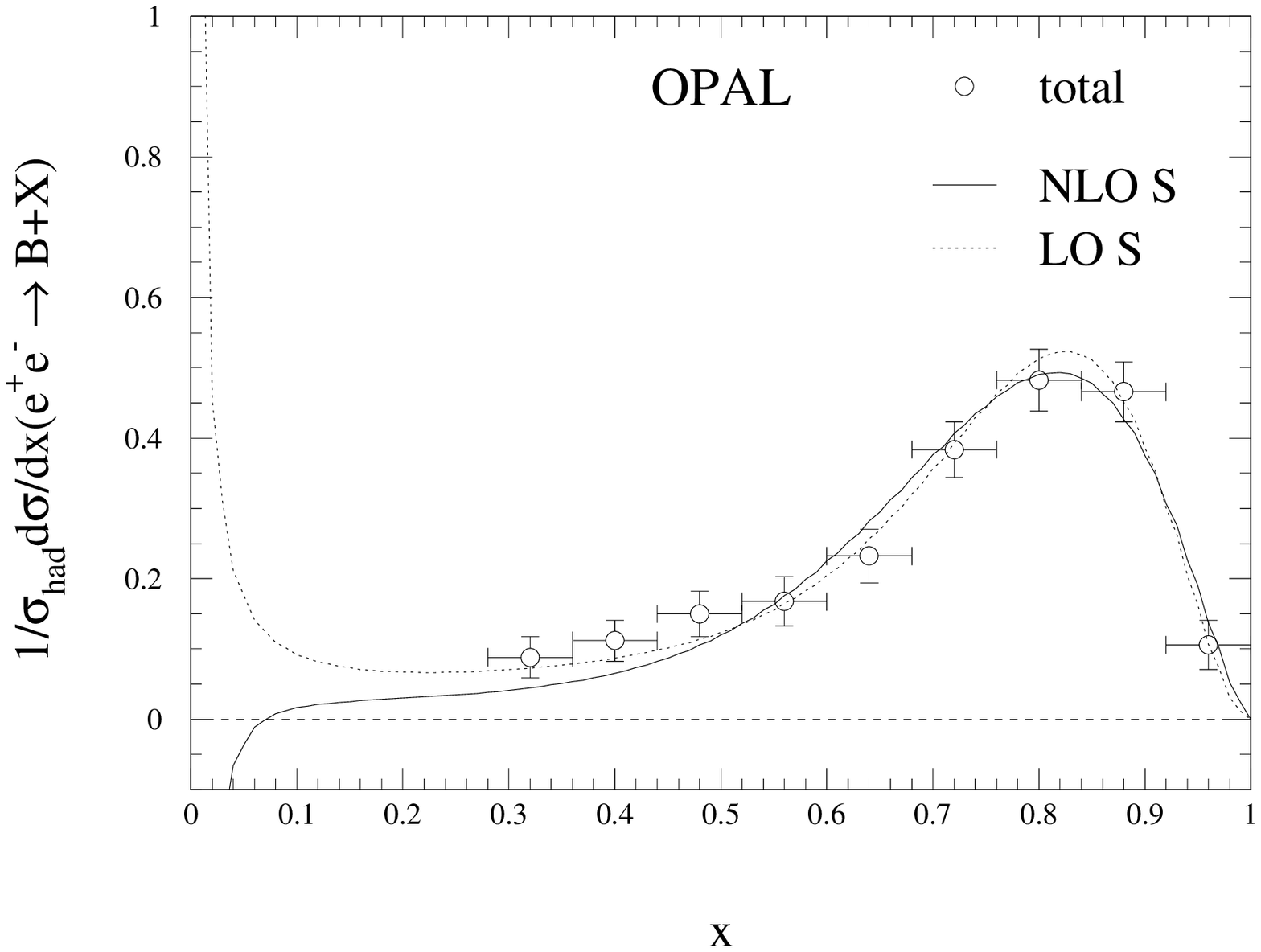,width=\textwidth}
\centerline{\Large\bf Fig.~1a}
\end{figure}

\newpage
\begin{figure}[ht]
\epsfig{figure=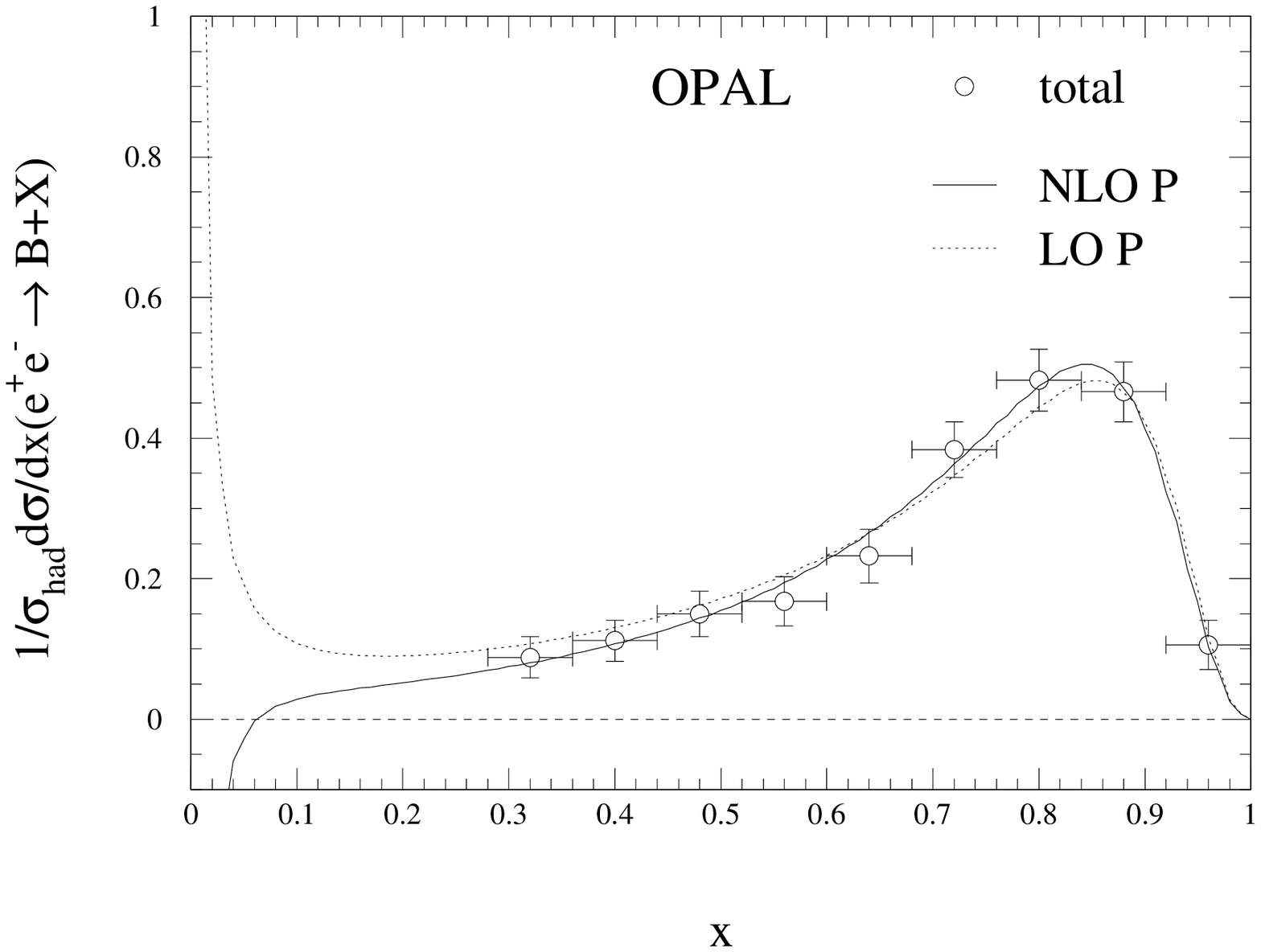,width=\textwidth}
\centerline{\Large\bf Fig.~1b}
\end{figure}

\newpage
\begin{figure}[ht]
\epsfig{figure=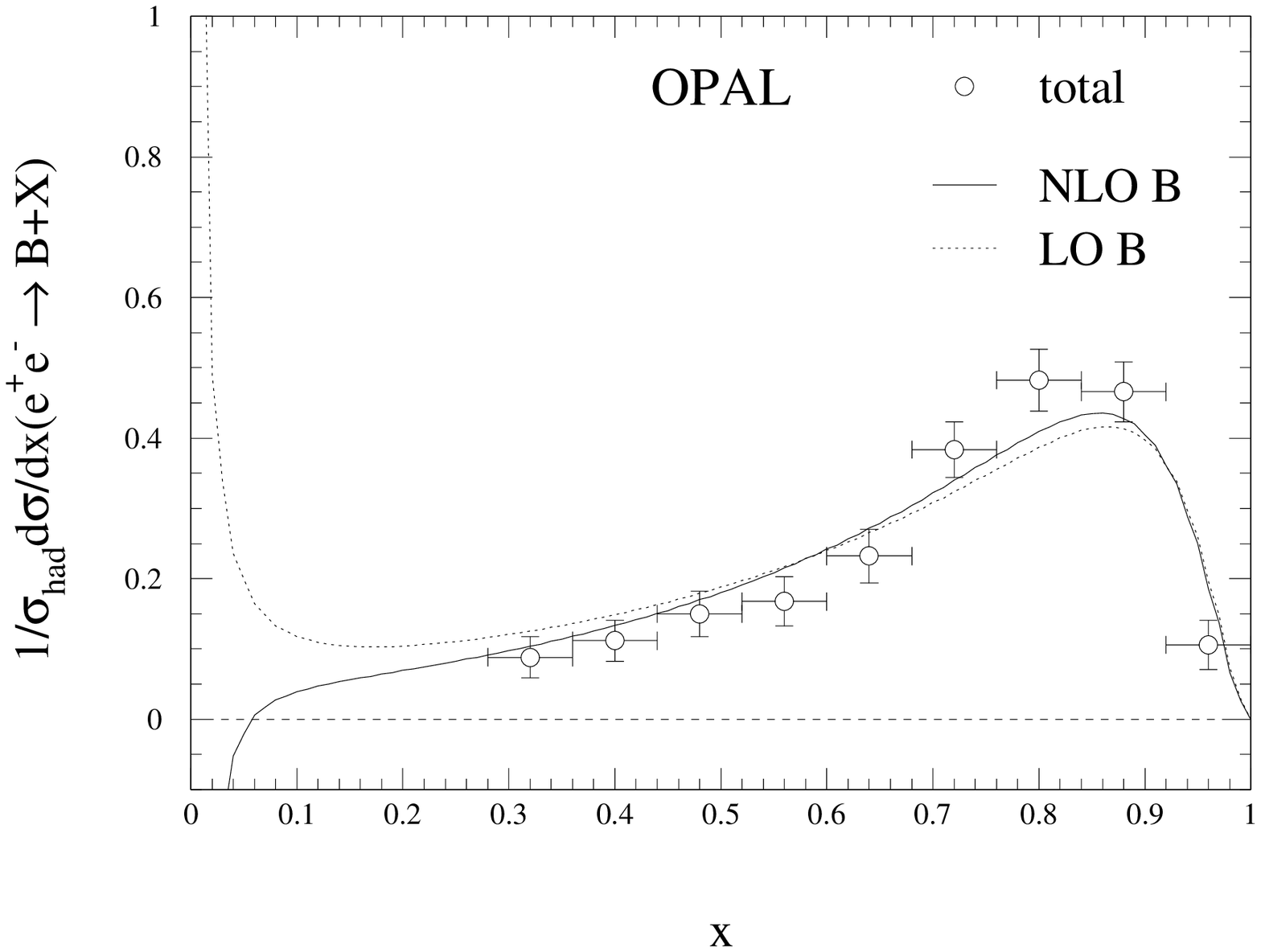,width=\textwidth}
\centerline{\Large\bf Fig.~1c}
\end{figure}

\newpage
\begin{figure}[ht]
\epsfig{figure=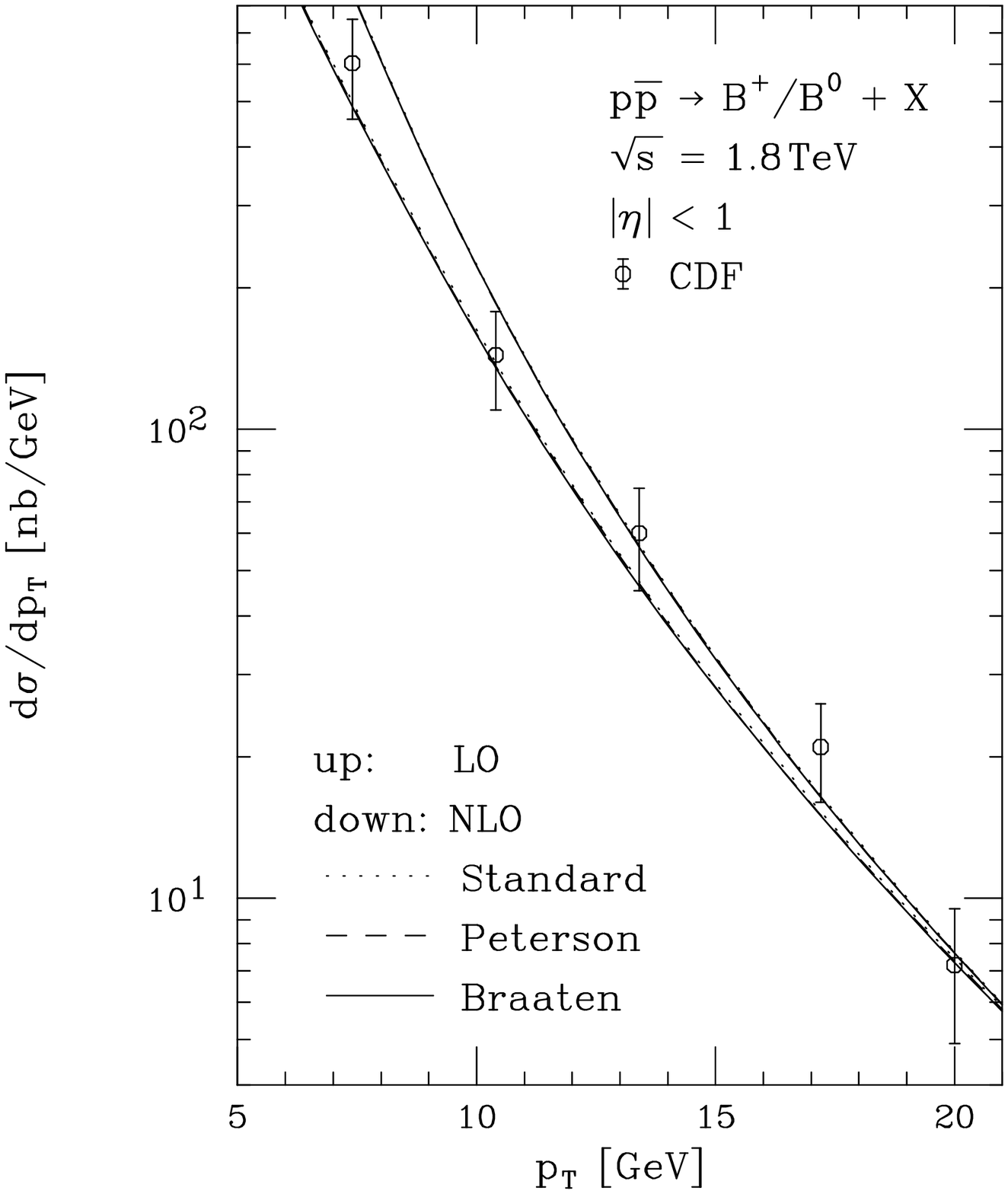,width=\textwidth}
\centerline{\Large\bf Fig.~2a}
\end{figure}

\newpage
\begin{figure}[ht]
\epsfig{figure=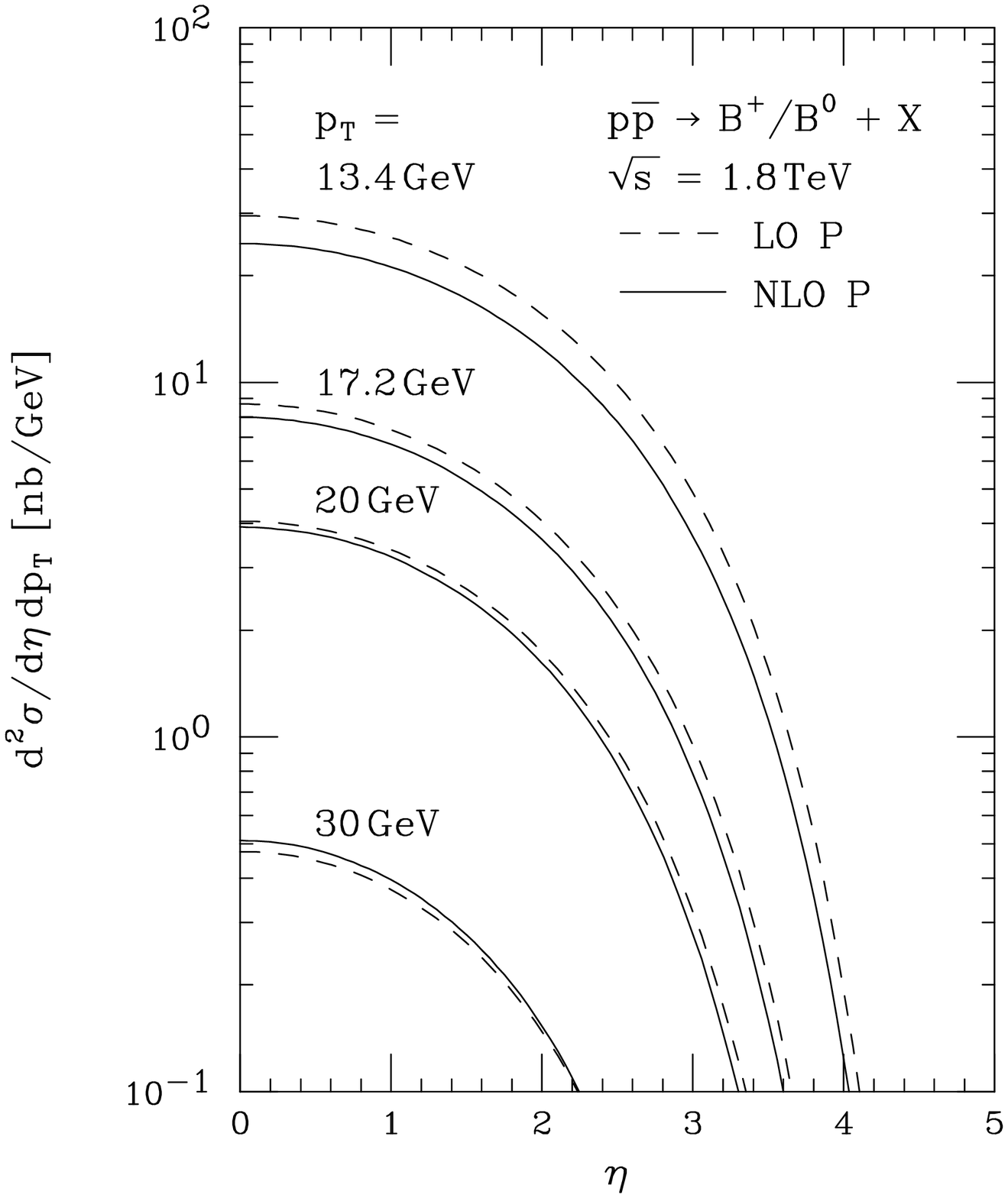,width=\textwidth}
\centerline{\Large\bf Fig.~2b}
\end{figure}

\newpage
\begin{figure}[ht]
\epsfig{figure=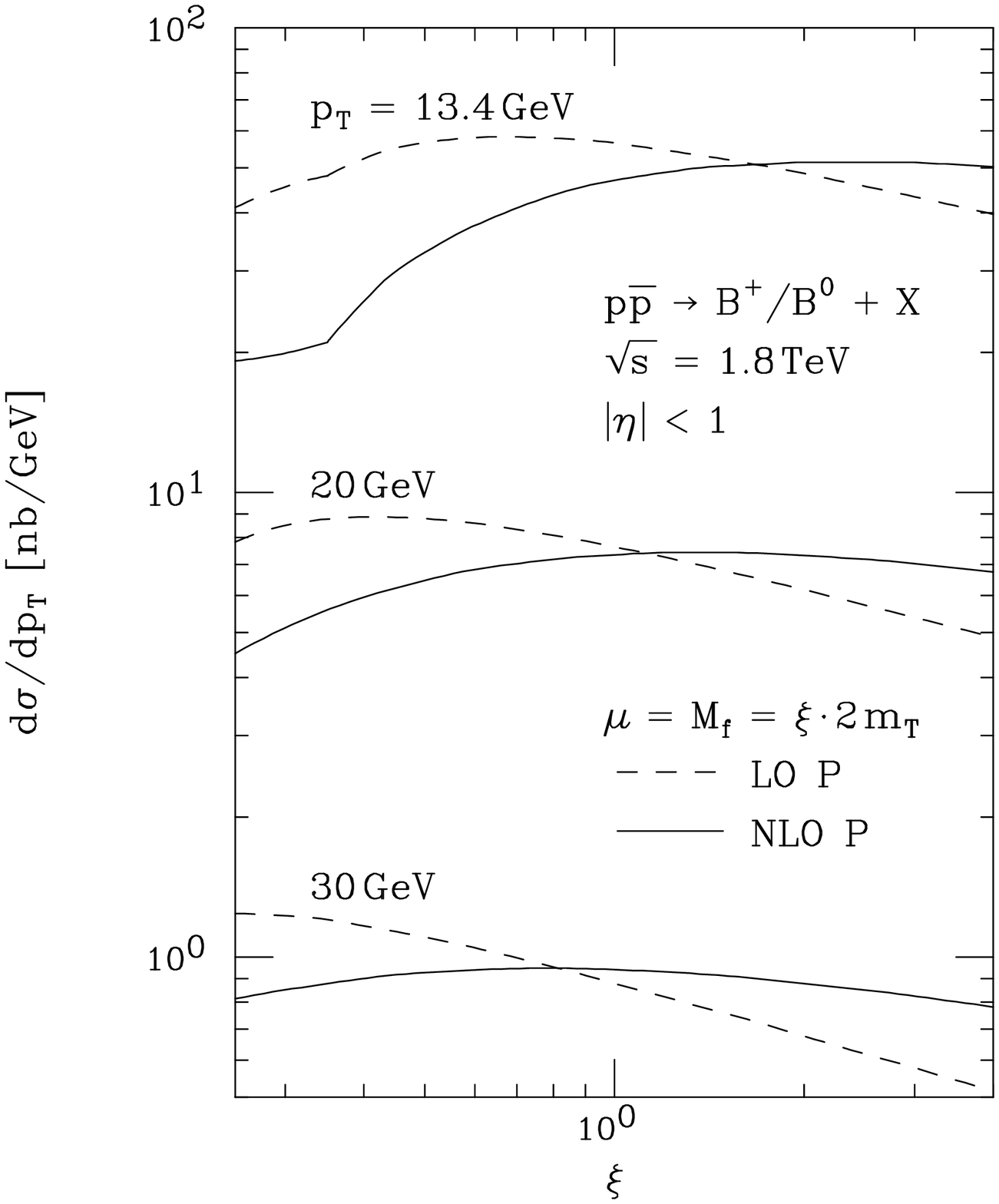,width=\textwidth}
\centerline{\Large\bf Fig.~3}
\end{figure}

\end{document}